# Noise Level Estimation for Overcomplete Dictionary Learning Based on Tight Asymptotic Bounds

Rui Chen, *Member, IEEE,* Changshui Yang, Huizhu Jia and Xiaodong Xie

*Abstract*—In this letter, we address the problem of estimating Gaussian noise level from the trained dictionaries in update stage. We first provide rigorous statistical analysis on the eigenvalue distributions of a sample covariance matrix. Then we propose an interval-bounded estimator for noise variance in high dimensional setting. To this end, an effective estimation method for noise level is devised based on the boundness and asymptotic behavior of noise eigenvalue spectrum. The estimation performance of our method has been guaranteed both theoretically and empirically. The analysis and experiment results have demonstrated that the proposed algorithm can reliably infer true noise levels, and outperforms the relevant existing methods.

## I. Introduction

THE dictionary learning is a matrix factorization problem that amounts to finding the linear combination of a given signal $\mathbf{Y} \in \mathbb{R}^{N \times M}$ with only a few atoms selected from columns of the dictionary $\mathbf{D} \in \mathbb{R}^{N \times K}$. In an overcomplete setting, the dictionary matrix $\mathbf{D}$ has more columns than rows $K > N$, and the corresponding coefficient matrix $\mathbf{X} \in \mathbb{R}^{K \times M}$ is assumed to be sparse. For most practical tasks in the presence of noise, we consider a contamination form of the measurement signal $\mathbf{Y} = \mathbf{DX} + \mathbf{w}$, where the elements of noise $\mathbf{w}$ are independent realizations from the Gaussian distribution $\mathcal{N}(0, \sigma_n^2)$. The basic dictionary learning problem is formulated as:

$$\min_{\mathbf{D},\mathbf{X}} \|\mathbf{Y} - \mathbf{DX}\|_F^2 \quad s.t. \quad \|\mathbf{x}_i\|_0 \le L \ \forall i \tag{1}$$

Therein, $L$ is the maximal number of non-zero elements in the coefficient vector $\mathbf{x}_i$. Starting with an initial dictionary, this minimization task can be solved by the popular alternating approaches such as the method of optimal directions (MOD) [1] and K-SVD [2]. The dictionary training on noisy samples can incorporate the denoising together into one iterative process. In general, the residual errors of learning process are determined by noise levels [3]. Noise incursion in a trained dictionary can affect the stability and accuracy of sparse representation [4]. So the performance of dictionary learning highly depends on the estimation accuracy of unknown noise level $\sigma_n^2$ when the noise characteristics of trained dictionaries are unavailable.

The main challenge of estimating the noise level lies in effectively distinguishing the signal from noise by exploiting sufficient prior information. The most existing methods have been developed to estimate the noise level from image signals based on specific image characteristics [5]-[8]. Generally, these works assume that a sufficient amount of homogeneous areas or self-similarity patches are contained in natural images. Thus empirical observations, singular value decomposition (SVD) or statistical properties can be applied on carefully selected patches. However, it is not suitable for estimating the noise level in dictionary update stage because only few atoms for sparse representation cannot guarantee the usual assumptions. To enable wider applications and less assumptions, more recent methods estimate the noise level based on principal component analysis (PCA) [9], [10]. These methods underestimate the noise level since they only take the smallest eigenvalue of block covariance matrix. Although later work [11] has made efforts to tackle these problems by spanning low dimensional subspace, the optimal estimation for true noise variance is still not achieved due to the inaccuracy of subspace segmentation. As for estimating the noise variance techniques, the scaled median absolute deviation of wavelet coefficients has been widely adopted [12]. Leveraging the results from random matrix theory (RMT), the median of sample eigenvalues is also used as an estimator of noise variance [13]. However, these estimators are no longer consistent and unbiased when the dictionary matrix has high dimensional structure.

To solve the aforementioned problems, we propose to accurately estimate noise variance by using exact eigenvalues of sample covariance matrix. A tight asymptotic bound for extreme eigenvalues is constructed to separate the subspaces between the signal and noise. For trained dictionaries with low-sample sizes and high dimensions, a bounded estimator provides a consistent inference on noise variance. The practical usefulness of our method is numerically illustrated.

## II. Tight Bound for Noise Eigenvalue Distribution

In this section, we analyze the asymptotical distribution of the ratio of extreme eigenvalues of a sample covariance matrix based on the limiting RTM law. Then a tight bound is derived.

Manuscript received December XX, 2017; revised XX, 2017; accepted XX, 2017. Date of publication XX, 2017; date of current version XX, 2017. This work was supported by Beijing major science and technology projects under Grant Z171100000117008. The associate editor coordinating the review of this manuscript and approving it for publication was Prof. XXXX.

R. chen is with the School of Microelectronics, Tianjin University, Tianjin 300072, China (e-mail: rchen@jdl.ac.cn).

C. Yang* (Corresponding author), H. Jia and X. Xie are with National Engineering Laboratory for Video Technology, Peking University, Beijing 100871, China (e-mail: {csyang, hzjia, donxie}@pku.edu.cn).

Color versions of one or more of the figures in this paper are available online at http://ieeexplore.ieee.org.

Digital Object Identifier 10.1109/LSP.2015.2448732

*A. Eigenvalue Subspaces of Sample Covariance Matrix*

We consider the sparse approximation of each observed sample $\mathbf{y}_i \in \mathbb{R}^N$ with $s$ prototype atoms selected from learned dictionary $\mathbf{D}$. With respect to the sparse model (1), we aim at estimating the noise level $\sigma_n^2$ for an elementary trained dictionary $\mathbf{D}_s$ containing a subset of the atoms $\{\mathbf{d}_i\}_{i=1}^s$. At each iterative step, the noise level $\sigma_n^2$ goes gradually to zero when updating towards the true dictionary [14]. The known noise variance is helpful to avoid noise incursion and determine the sample size, the sparsity degree and even the performance of the true underlying dictionary [15]. To derive the relationship between the eigenvalues and noise level, we first construct the sample covariance matrix of dictionary $\mathbf{D}_s$ as follows:

$$\Sigma_S = \frac{1}{s-1}\sum_{i=1}^{s}(\mathbf{d}_i - \bar{\mathbf{d}})(\mathbf{d}_i - \bar{\mathbf{d}})^\mathrm{T}, \quad \bar{\mathbf{d}} = \frac{1}{s}\sum_{i=1}^{s}\mathbf{d}_i \qquad (2)$$

According to (2), the square matrix $\Sigma_S$ has $N$ dimensions with the sparse condition $N \gg s$. Based on the symmetric property, this matrix is decomposed into the product of three matrices: an orthogonal matrix $\mathbf{U}$, a diagonal matrix and a transpose matrix $\mathbf{U}^\mathrm{T}$, which can be selected by satisfying $\mathbf{U}^\mathrm{T}\mathbf{U} = \mathbf{I}$. Here, this transform process is written as:

$$\mathbf{U}^\mathrm{T}\Sigma_S\mathbf{U} = \mathbf{diag}(\lambda_1, ..., \lambda_m, \lambda_{m+1}, ..., \lambda_N) \qquad (3)$$

Given $\lambda_1 \geq \lambda_2 \geq ... \geq \lambda_N$, we exploit the eigenvalue subspaces to enable the separation of atoms from noise. To be more specific, we divide the eigenvalues into two sets $\mathbf{S} = \mathbf{S}_1 \cup \mathbf{S}_2$ by finding the appropriate bound in a spiked population model [16]. Most structures of an atom lie in low-dimension subspace and thus the leading eigenvalues in set $\mathbf{S}_1 = \{\lambda_i\}_{i=1}^{m}$ are mainly contributed by atom itself. The redundant-dimension subspace $\mathbf{S}_2 = \{\lambda_i\}_{i=m+1}^{N}$ is dominated by the noise. Because the atoms contribute very little to this later portion, we take all the eigenvalues of $\mathbf{S}_2$ into consideration to estimate the noise variance while eliminating the influence of trained atoms. Moreover, the random variables $\{\lambda_i\}_{i=m+1}^{N}$ can be considered as the eigenvalues of pure noise covariance matrix $\Sigma_\mathbf{w}$, whose dimensions are $N$.

*B. Asymptotic Bound for Noise Eigenvalues*

Suppose the sample matrix $\Sigma_\mathbf{w}$ has the form $(s-1)\Sigma_\mathbf{w} = \mathbf{HH}^\mathrm{T}$, where the sample entries of $\mathbf{H}$ are independently generated from the distribution $\mathcal{N}(0, \sigma_n^2)$. Then the real matrix $\mathbf{M} = \mathbf{HH}^\mathrm{T}$ follows a standard Wishart distribution [17]. The ordered eigenvalues of $\mathbf{M}$ are denoted by $\bar{\lambda}_{\max}(\mathbf{M}) \geq \cdots \geq \bar{\lambda}_{\min}(\mathbf{M})$. In the high dimensional situation: $N/s \to \gamma \in [0, \infty)$ as $s$ fixed and $N \to \infty$, the Tracy-Widom law gives the limiting distribution of the largest eigenvalue of the large random matrix $\mathbf{M}$ [18]. Then we have the following asymptotic expression:

$$\Pr\left\{\frac{\bar{\lambda}_{\max}/\sigma_n^2 - \mu}{\xi} \leq z\right\} \to F_{\mathrm{TW1}}(z) \qquad (4)$$

where $F_{\mathrm{TW1}}(z)$ indicates the cumulative distribution function with respect to the Tracy-Widom random variable. In order to improve both the approximation accuracy and convergence rate, even only with few atom samples, we need choose the suitable centering and scaling parameters $\mu, \xi$ [19]. By the comparison between different values, such parameters are defined as

$$\begin{cases} \mu = \left(\sqrt{s-1/2} + \sqrt{N-1/2}\right)^2 \\ \xi = \left(\sqrt{s-1/2} + \sqrt{N-1/2}\right)\left(\frac{1}{\sqrt{s-1/2}} + \frac{1}{\sqrt{N-1/2}}\right)^{1/3} \end{cases} \qquad (5)$$

The empirical distribution of the eigenvalues of the large sample matrix converges almost surely to the Marcenko-Pastur distribution on a finite support [20]. Based on the generalized result in [21], when $N \to \infty$ and $\gamma \in [0, \infty)$, with probability one, we derive limiting value of the smallest eigenvalue as

$$\bar{\lambda}_{\min}/\sigma_n^2 \to \left(1 - \sqrt{\gamma}\right)^2 \qquad (6)$$

According to the asymptotic distributions described in the theorems (4) and (6), we further quantify the distribution of the ratio of the maximum eigenvalue to minimum eigenvalue in order to detect the noise eigenvalues. Let $T_1$ be a detection threshold. Then we find $T_1$ by the following expression:

$$\Pr\left\{\frac{\bar{\lambda}_{\max}}{\bar{\lambda}_{\min}} \leq T_1\right\} = \Pr\left\{\frac{\bar{\lambda}_{\max}}{\sigma_n^2} \leq T_1 \cdot \frac{\bar{\lambda}_{\min}}{\sigma_n^2}\right\} \approx \Pr\left\{\frac{\bar{\lambda}_{\max}}{\sigma_n^2} \leq T_1 \cdot \left(1 - \sqrt{N/s}\right)^2\right\}$$
$$= \Pr\left\{\frac{\bar{\lambda}_{\max}/\sigma_n^2 - \mu}{\xi} \leq \frac{T_1 \cdot \left(1 - \sqrt{N/s}\right)^2 - \mu}{\xi}\right\} \approx F_{\mathrm{TW1}}\left\{\frac{T_1 \cdot \left(1 - \sqrt{N/s}\right)^2 - \mu}{\xi}\right\} \qquad (7)$$

Note that there is no closed-form expression for the function $F_{\mathrm{TW1}}$. Fortunately, the values of $F_{\mathrm{TW1}}$ and the inverse $F_{\mathrm{TW1}}^{-1}$ can be numerically computed at certain percentile points [16]. For a required detection probability $\alpha_1$, this leads to

$$\frac{T_1 \cdot \left(1 - \sqrt{N/s}\right)^2 - \mu}{\xi} = F_{\mathrm{TW1}}^{-1}(\alpha_1) \qquad (8)$$

Plugging the definitions of $\mu$ and $\xi$ into the Eq. (8), we finally obtain the threshold

$$T_1 = \frac{s\left(\sqrt{s-1/2} + \sqrt{N-1/2}\right)^2}{\left(\sqrt{s} - \sqrt{N}\right)^2} \cdot \left(\frac{\left(\sqrt{s-1/2} + \sqrt{N-1/2}\right)^{-2/3}}{(s-1/2)^{1/6}(N-1/2)^{1/6}} \cdot F_{\mathrm{TW1}}^{-1}(\alpha_1) + 1\right) \qquad (9)$$

When the detection threshold $T_1$ is known in the given probability, it means that an asymptotic upper bound can also be obtained for determining the noise eigenvalues of the matrix $\Sigma_\mathbf{w}$ because the equality $\lambda_{m+1}/\lambda_N = \bar{\lambda}_{\max}/\bar{\lambda}_{\min}$ holds. In general, the noise eigenvalues in the set $\mathbf{S}_2$ surround the true noise variance as it follows the Gaussian distribution. The estimated largest eigenvalue $\lambda_{m+1}$ should be no less than $\sigma_n^2$. The known smallest eigenvalue $\lambda_N$ is no more than $\sigma_n^2$ by the theoretical analysis [11]. The location and value of $\lambda_{m+1}$ in $\mathbf{S}$ are obtained by checking the bound $\lambda_{m+1} \leq T_1 \cdot \lambda_N$ with high probability $\alpha_1$. In addition, $\lambda_1$ cannot be selected as noise eigenvalue $\lambda_{m+1}$.

## III. Noise Variance Estimation Algorithm

### A. Bounded Estimator for Noise Variance

Without requiring the knowledge of signal, the threshold $T_1$ can provide good detection performance for finite $s$, $N$ even when the ratio $N/s$ is not too large. Based on this result, more accurate estimation can be obtained by averaging all elements in $\mathbf{S}_2$. Hence, the maximum likelihood estimator of $\sigma_n^2$ is

$$\hat{\sigma}_n^2 = \frac{1}{N-m} \sum_{j=m+1}^{N} \lambda_j \quad (10)$$

In the low dimensional setting where $N$ is relatively small compared with $s$, the estimator $\hat{\sigma}_n^2$ is consistent and unbiased as $s \to \infty$. It follows asymptotically normal distribution as

$$\sqrt{s}(\hat{\sigma}_n^2 - \sigma_n^2) \to \mathcal{N}(0, t^2), \quad t^2 = \frac{2\sigma_n^4}{N-m} \quad (11)$$

When $N$ is large with respect to the sample size $s$, the sample covariance matrix shows significant deviations from the underlying population covariance matrix. In this context, the estimator $\hat{\sigma}_n^2$ might have a negative bias, which leads to overestimation of true noise variance [22], [23]. We investigate the distribution of another eigenvalue ratio. Namely, the ratio of the maximum eigenvalue to the trace of the eigenvalues is

$$U = \frac{\lambda_{m+1}}{1/(N-m) \cdot \text{tr}(\Sigma_\mathbf{w})} = \frac{\lambda_{m+1}}{1/(N-m) \cdot \sum_{j=m+1}^{N} \lambda_j} \quad (12)$$

According to the result in (4), the ratio $U$ also follows a Tracy-Widom distribution as both $N$, $s \to \infty$. The denominator in the definition of $U$ is distributed as an independent $\sigma_n^2 \chi_N^2/N$ random variable, and thus has $\mathrm{E}(\hat{\sigma}_n^2) = \sigma_n^2$ and $\mathrm{Var}(\hat{\sigma}_n^2) = 2\sigma_n^4/(N \cdot s)$. It is easy to show that replacing $\sigma_n^2$ by $\hat{\sigma}_n^2$ results in the same limiting distribution in (4). Then we have

$$\Pr\left\{\frac{\lambda_{m+1}/\hat{\sigma}_n^2 - \mu}{\xi} \le z\right\} \to F_{\mathrm{TW1}}(z) \quad (13)$$

Unfortunately, the asymptotic approximation present in (13) is inaccurate for small and even moderate values of $N$ [24]. This approximation is not a proper distribution function. The simulation observations imply that the major factor contributing to the poor approximation is the asymptotic error caused by the constant $\xi$ [24]. Therefore, a more accurate estimate for the standard deviation of $\lambda_{m+1}/\hat{\sigma}_n^2$ will provide a significant improvement. For finite samples, we have

$$\mathrm{E}\left(\frac{\lambda_{m+1}}{\sigma_n^2}\right) = \mu, \quad \mathrm{E}\left(\frac{\lambda_{m+1}^4}{\sigma_n^4}\right) = \mu^2 + \xi^2 \quad (14)$$

Using these asymptotic results, we get the corrected deviation

$$\xi' = \sqrt{\frac{N \cdot s}{2 + N \cdot s}\left(\xi^2 - \frac{2}{N \cdot s}\mu^2\right)} \quad (15)$$

Note that this formula in (15) has corrected the overestimation in the high dimensional setting. Thus the better approximation for the probabilities of the ratio is

$$\Pr\left\{\frac{\lambda_{m+1}/\hat{\sigma}_n^2 - \mu}{\xi'} \ge z\right\} \approx 1 - F_{\mathrm{TW1}}(z) \quad (16)$$

The determination of the distribution for the ratio $U$ is devoted to the correction of the variance estimator. In order to complete the detection of the large deviations of the initial estimator $\hat{\sigma}_n^2$, we provide a procedure to set the threshold $T_2$. Based on the result in (16), an approximate expression for the overestimation probability is given by

$$\Pr\left\{\frac{\hat{\sigma}_n^2}{\lambda_{m+1}} \le T_2\right\} = \Pr\left\{\frac{\lambda_{m+1}/\hat{\sigma}_n^2 - \mu}{\xi'} \ge \frac{1/T_2 - \mu}{\xi'}\right\} \approx 1 - F_{\mathrm{TW1}}\left(\frac{1/T_2 - \mu}{\xi'}\right) \quad (17)$$

Hence, for a desired probability level $\alpha_2$, the above equation can be numerically inverted to find the decision threshold. After some simplified manipulations, we obtain

$$T_2 = \frac{1}{\xi' \cdot F_{\mathrm{TW1}}^{-1}(1-\alpha_2) + \mu} \quad (18)$$

Asymptotically, the spike eigenvalue $\lambda_{m+1}$ converges to the right edge of the support $\sigma_n^2(1+\sqrt{N/s})$ as $N$, $s$ go to infinity. According to the expression in (18), this function turns out to have a simple approximation $T_2 = 1/\mu$ in the high probability case. Then the upper bound $T_2 \cdot \lambda_{m+1}$ for the known $\hat{\sigma}_n^2$ yields a bias estimation. Finally, the following expectation holds true:

$$\mathrm{E}\left(\frac{\mu \cdot T_2 \cdot \lambda_{m+1}}{1 + \sqrt{N/s}}\right) \approx \sigma_n^2 \ll \hat{\sigma}_n^2 \quad (19)$$

By analyzing the statistical result in (19), the correction for $T_2 \cdot \lambda_{m+1}$ can be approximated as the better estimator than $\hat{\sigma}_n^2$ because this bias-corrected estimator is closer to the true variance under the high dimensional conditions. If $\hat{\sigma}_n^2$ can satisfy the requirement of no excess of the bound $T_2 \cdot \lambda_{m+1}$, the sample eigenvalues are consistent estimates of their population counterparts. Hence, the optimal estimator is given by

$$\hat{\sigma}_*^2 = \min\left\{\hat{\sigma}_n^2, \frac{\mu \cdot T_2 \cdot \lambda_{m+1}}{1 + \sqrt{N/s}}\right\} \quad (20)$$

### B. Implementation

Based on the construction of two thresholds, we propose the noise estimation method for dictionary learning as follows:

**Step 1.** Compute the eigenvalues $\{\lambda_i\}_{i=1}^{N}$ of the sample covariance matrix $\Sigma_s$, and order $\lambda_1 \ge \lambda_2 \ge ... \ge \lambda_N$.

**Step 2.** Set the probability levels $\alpha_1$ and $\alpha_2$.

**Step 3.** Compute two thresholds $T_1$ and $T_2$.

**Step 4.** Obtain the location $m+1$ of noise eigenvalues and the value of $\lambda_{m+1}$ by checking whether $\lambda_{m+1} \le T_1 \cdot \lambda_N$ is true.

**Step 5.** Compute the initial estimator $\hat{\sigma}_n^2$.

**Step 6.** Compare two estimators in (20) and select the minimum as the optimal estimation of $\sigma_n^2$.

## IV. Numerical Experiments

The proposed estimation method is evaluated on the images of size $768 \times 512$ from Kodak database [7]. The subjective experiment is to compare our method with three state-of-the-art estimation methods by Liu et al. in [8], Pyatykh *et al.* in [9] and Chen *et al.* in [11]. The testing images including **Woman** and **House** are added to the independent white Gaussian noise with deviation level 10 and 30, respectively. We set the probabilities $\alpha_1, \alpha_2 = 0.97$ and choose $N = 256$ and $s = 3$. In general, a higher noise estimation accuracy leads to a higher denoising quality. We use the K-SVD method to denoise the images [3]. Figs. (1) and (2) show the results using our method outperform other competitors. Moreover, our peak signal-to-noise ratios (PSNRs) are nearest to true values, 32.03 dB and 27.01 dB, respectively.

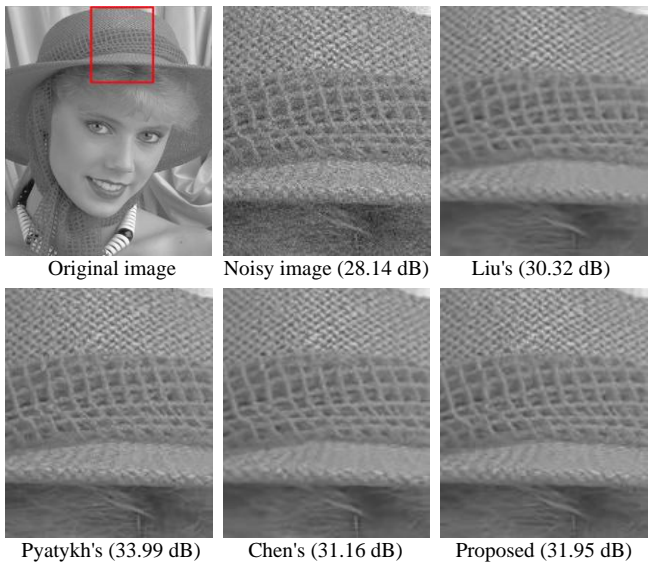

| Original image | Noisy image (28.14 dB) | Liu's (30.32 dB) |
| Pyatykh's (33.99 dB) | Chen's (31.16 dB) | Proposed (31.95 dB) |

Fig. 1. Denoising results on the **Woman** image using K-SVD.

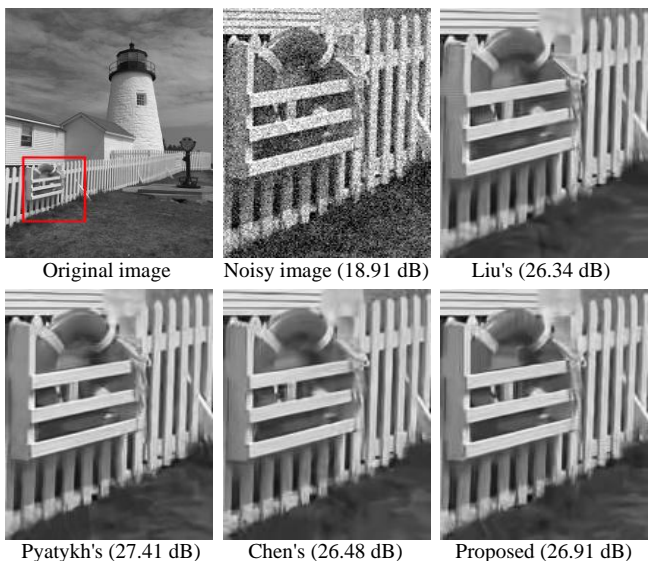

| Original image | Noisy image (18.91 dB) | Liu's (26.34 dB) |
| Pyatykh's (27.41 dB) | Chen's (26.48 dB) | Proposed (26.91 dB) |

Fig. 2. Denoising results on the **House** image using K-SVD.

To quantitatively evaluate the accuracy of noise estimation, the average of standard deviations, mean square error (MSE), mean absolute difference (MAD) are computed by randomly selecting 1000 image patches from the testing images. The results shown in Table I indicate that the proposed method is more accurate and stable. Next, we compare our estimator $\hat{\sigma}_*^2$ with $\hat{\sigma}_n^2$ and other two existing estimators in the literature. The simulated realization of a sample covariance matrix is followed a Gaussian distribution with different variances. As presented in Table II, the performance of $\hat{\sigma}_*^2$ is invariably better than other estimators. To test robustness of our method, we further obtain the empirical probabilities of estimated eigenvalues at typical confidence levels. Fig.3 illustrates that two asymptotic bounds can achieve very high success probabilities.

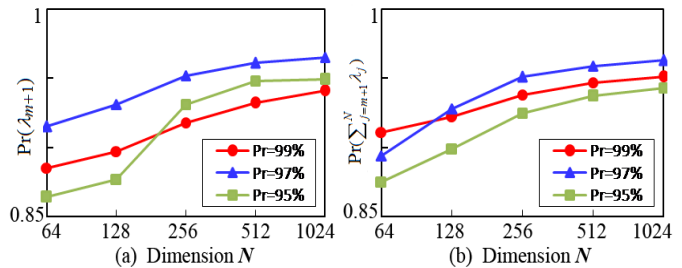

Fig. 3. Empirical probabilities of exact noise eigenvalue estimation.

TABLE I
ESTIMATION RESULTS OF DIFFERENT METHODS (BEST RESULTS HIGHLIGHTED)

| $\sigma_n$ | Liu's [8] | Pyatykh's [9] | Chen's [11] | Proposed |
|---|---|---|---|---|
| 1 | 2.21 | 1.26 | 0.64 | **1.17** |
| 5 | 7.35 | 3.82 | 5.38 | **5.24** |
| 10 | 13.96 | 7.16 | 11.84 | **10.19** |
| 15 | 16.75 | 13.93 | 15.92 | **15.11** |
| 20 | 20.96 | 18.74 | 20.54 | **19.92** |
| 25 | 26.64 | 23.26 | 24.39 | **25.07** |
| 30 | 32.34 | 27.28 | 31.95 | **30.05** |
| MAD | 2.03 | 1.58 | 0.94 | **0.13** |
| MSE | 3.36 | 2.57 | 1.21 | **0.02** |

TABLE II
ESTIMATION RESULTS OF FOUR ESTIMATORS (BEST RESULTS HIGHLIGHTED)

| $\sigma_n$ | $\hat{\sigma}_{\text{median}}$ [23] | $\hat{\sigma}_{\text{US}}$ [13] | $\hat{\sigma}_n$ | $\hat{\sigma}_*$ |
|---|---|---|---|---|
| 1 | 1.28 | 1.94 | 1.15 | **1.04** |
| 5 | 4.59 | 5.23 | 6.27 | **5.12** |
| 10 | 8.67 | 11.24 | **9.92** | 9.92 |
| 15 | 15.27 | 14.09 | 16.08 | **14.97** |
| 20 | 20.73 | 19.24 | 20.97 | **20.08** |
| 25 | 25.78 | 25.93 | 26.25 | **25.13** |
| 30 | 30.45 | 30.26 | 31.19 | **30.03** |
| MAD | 0.61 | 0.75 | 0.86 | **0.07** |
| MSE | 0.69 | 1.63 | 1.16 | **0.02** |

## V. Conclusions

In this letter, we have shown how to infer the noise level from a trained dictionary. The eigen-spaces of the signal and noise are transformed and separated well by determining the eigen-spectrum interval. In addition, the developed estimator can effectively eliminate the estimation bias of noise variance in high dimensional context. Our noise estimation technique has low computational complexity. The experimental results have demonstrated that our method outperforms the relevant existing methods over a wide range of noise level conditions.